\documentclass[aps,pre,showpacs,amssymb,twocolumn]{revtex4}
\usepackage[dvips]{graphicx}
\usepackage{bm}
\usepackage[bookmarks]{hyperref}

\begin{document}

\title{Criticality in a Vlasov--Poisson system -- a fermionic universality class}

\author{A.~V. Ivanov}
\author{S.~V. Vladimirov}
\author{P.~A. Robinson}

\affiliation{School of Physics, The University of Sydney, NSW
2006, Sydney, Australia}

\begin{abstract}

A model Vlasov--Poisson system is simulated close the point of
marginal stability, thus assuming only the wave-particle resonant
interactions are responsible for saturation, and shown to obey the
power--law scaling of a second-order phase transition. The set of
critical exponents analogous to those of the Ising universality
class is calculated and shown to obey the Widom and Rushbrooke
scaling and Josephson's hyperscaling relations at the formal
dimensionality $d=5$ below the critical point at nonzero order
parameter. However, the two-point correlation function does not
correspond to the propagator of Euclidean quantum field theory,
which is the Gaussian model for the Ising universality class.
Instead it corresponds to the propagator for the fermionic {\it
vector} field and to the {\it upper critical dimensionality}
$d_c=2$. This suggests criticality of collisionless Vlasov-Poisson
systems as representative of the {\it universality class} of
critical phenomena of {\it a fermionic} quantum field description.

\end{abstract}


\maketitle

\section{Introduction}{\label{intro}}

The remarkable property of critical phenomena is the universal
scaling appearing in vast variety of systems; e.g., magnets and
gases follow simple power laws for the order parameter, specific
heat capacity, susceptibility, compressibility, etc. \cite{st}. In
thermodynamic systems, phase transitions take place at a critical
temperature $T_{cr}$ when the coefficients that characterize the
linear response of the system to external perturbations {\it
diverge}. Then long--range order appears, causing a transition to
a new phase due to collective behavior of the entire system
\cite{land}.

The condition for nonlinear saturation in the test case of the
bump--on--tail instability in plasmas \cite{bump} is
\begin{equation}\label{scalb}
    \omega_b \approx 3.2 \gamma_L
\end{equation}
\cite{fried,levin_ea}, where $\gamma_L$ is the linear growth rate
of a weakly unstable Langmuir wave according to the Landau theory
\cite{land1}, $\omega_b=(eEk/m)^{1/2}$ is the frequency of
oscillations for particles trapped by the wave. These trapped
particles generate a long--range order of the wavelength $k$, the
saturated amplitude $E$ can be considered as the order parameter,
and the condition of saturation can be rewritten as a power law,
typical for the second-order phase transitions, $E\sim
\gamma_L^\beta$.

However, unlike thermodynamics this scaling contains the
nonthermal control parameter $\gamma_L$, which is determined by
the slope of the distribution function $\partial f_0/\partial v$
at the phase speed $v_r=\omega_{pe}/k$ of the perturbation near
the electron plasma frequency $\omega_{pe}$. [For thermodynamic
systems like magnets for which the magnetization $M$ below the
Curie point $T_{cr}$ is the order parameter, the scaling is $M
\propto \epsilon^\beta$, where $\epsilon=(T-T_{cr})/T_{cr}$ at
$T<T_{cr}$.] Another difference is the critical exponent itself --
relation (\ref{scalb}) predicts the very unusual exponent
$\beta=2$; in contrast, the hydrodynamic Hopf bifurcation
\cite{hopf}, also described by the same scaling between the
saturated amplitude and the growth rate, has the mean--field
critical exponent $\beta=1/2$.

An analysis \cite{latora} assuming thermalization in a
Vlasov--Poisson plasma or gravitating system leads to a critical
exponent $\beta<1$, and the exponent $\beta=1/2$ has also been
hypothesized for the bump-on-tail instability in \cite{rozen}.
However, detailed center-manifold analysis, which establishes the
normal form for a weakly unstable perturbation in one-component
collisionless Vlasov-Poisson system, confirms $\beta=2$
\cite{crawf}. The exponent $\beta=2$ is also confirmed numerically
\cite{denavit,candy}.

The striking discrepancy between these exponents can be better
understood if we consider the structure of the phase space
corresponding to these cases. The exponents $\beta=1/2$
\cite{hopf} and $\beta<1$ \cite{latora} correspond either to
saturation of a strongly {\it dissipative} instability or to a
thermalized system. In both these cases the distribution function
can be factorized as $f(q,p)=y(q)g(p)$, where $g(p)$ can be
assumed to be Gaussian, and the system is described by its
momenta. The exponent $\beta=2$ corresponds to saturation due to
nonlinear wave-particle interactions in a weakly unstable {\it
collisionless} system, where correlations between coordinates and
impulses are not destroyed by dissipative processes, so the
description cannot be reduced to moments of the distribution.

More formally, a dissipative and/or thermalized system is
represented by a {\it discrete} set of momenta of the distribution
function $f(x,v,t)$, which depend only on the coordinate $x$, but
not on the velocity $v$, $M=\{\rho,\bar v, T \}$, where $\rho$,
$\bar v$, and $T$ are the local density,  the velocity, and the
temperature, respectively. The evolution is a flow $g_t$ which
maps $M$ onto itself, $g_t\!\! :M\to M$. In fact, in a
neighborhood of the threshold $\gamma_L=0$ the evolution $g_t\!\!
:M\to M$ can be reduced to a normal form, which maps only the
order parameter, $g^\prime_t\!\! :n\to n$, where $n=\mathbb{R}^0$
(or $\mathbb{C}^0$, where $\mathbb{C}$ is the set of complex
numbers), and therefore the evolution is a {\it trajectory}
$n=n(t)$ or in other words the set $Y=\mathbb{R^+} \times
\mathbb{R}^0$. The phase space of a one-dimensional collisionless
system is a {\it continuous} set $H=\mathbb{R} \times \mathbb{R}$
and evolution can be represented as the flow $w_t \!\! : H \to H$.
(For periodic boundary conditions the phase space is isomorphic to
a cylinder $C=\mathbb{T} \times \mathbb{R}$, where $\mathbb{T}$ is
isomorphic to a circle.)  The sets $n$ and $H$ (or $C$) have
different {\it dimensionality}, and therefore renormalization of
collisionless system in a vicinity of threshold -- i.e.,
transformation of the set $H$ {\it and} the mapping $w_t$ --
involves one or more additional dimension. Further it is shown
below, that scaling transformations close to the threshold are
interrelated with the additional velocity coordinate, which
disappears in hydrodynamic description because of integration of
the distribution function $f(x,v,t)$ over $v$.

From the theory of critical phenomena it is known that
dimensionality $d$ is inseparable part of the threshold
description -- along with the critical exponents (e.g.,
\cite{amit}). Besides $\beta$, other critical exponents: $\alpha$,
$\gamma$, $\delta$, $\nu$, and $\eta$, describe the following
scalings of the Ising universality class: (i) the specific heat
capacity scales as
\begin{equation}
    C=\frac{\delta Q}{dT}\propto|\epsilon|^{-\alpha}~;
\end{equation}
(ii) the susceptibility as
\begin{equation}
    \chi=\left(\frac{\partial M}{\partial B}\right)_{B\to 0}
    \propto |\epsilon|^{-\gamma}~;
\end{equation}
(iii) the response $M$ at $\epsilon=0$ as
\begin{equation}
    M\propto B^{1/\delta}~;
\end{equation}
(iv) the correlation length as
\begin{equation}
    \xi\sim|\epsilon|^{-\nu}~;
\end{equation}
and (v) the two-point correlation function as
\begin{equation}
    G(r)\sim \frac{e^{-r/\xi}}{r^{d-2+\eta}}~.
\end{equation}

These exponents are not independent, but are interrelated via
scaling laws, e.g., the Widom equality
\begin{equation}\label{wideq}
    \gamma=\beta(\delta-1)
\end{equation}
\cite{widom}. These scaling laws also include \emph{hyperscaling}
laws such as Josephson's law,
\begin{equation}
    \label{josephl}
    \nu d=2-\alpha~,
\end{equation}
\cite{joseph} which involves the dimensionality $d$ along with the
exponents.

For thermodynamics the mean--field exponents are of the
Landau-Weiss set, $\alpha=0$, $\beta=1/2$, $\gamma=1$, $\delta=3$,
$\nu=1/2$, and $\eta=0$, and the scaling laws hold at the formal
dimensionality $d=4$. However, the possibilities of critical
phenomena are not exhausted by the Ising universality class -- the
percolation critical exponents \cite{staufer}, which describe
another vast class of critical phenomena, are different from those
in thermodynamics and scaling laws hold at a different
dimensionality. In particular, for the Bethe lattice (or Cayley
tree) \cite{domb} Josephson's law holds at dimensionality $d=6$.
Despite the description being the same, this difference separates
the cases into different \emph{universality classes} with
different \emph{upper critical dimensions}: $d_c=4$ for the Ising
universality class \cite{gold} and $d=6$ for percolation.

For a collisionless gravitating system, where the saturation
mechanism is the same as for the bump-on-tail instability in
plasmas, the critical exponent $\beta=1.907 \pm 0.006$, and the
critical exponents $\gamma=1.075 \pm 0.05$, $\delta=1.544 \pm
0.002$ can be determined analogously to thermodynamics and
calculated from the response to an external pump \cite{iva}. These
exponents are very different from the thermodynamic set, but
nevertheless satisfy the Widom equality, thus suggesting the
validity of scaling laws. Josephson's law also holds, but at a
rather surprising dimensionality which is the fractal one $d
\approx 4.68$ \cite{iva}. At the same time, the processes
resulting in $\beta \approx 1.9$ differ {\it qualitatively} from
those resulting in $\beta=2$, similar to thermodynamics where
spatial fluctuations of the order parameter, neglected in mean
field theories, result in $\beta \approx 0.33$, therefore
suggesting other universality classes were not completely ruled
out. These could be the wave-wave interactions, responsible for
the strong turbulence in plasma \cite{rob}), which are next in
dynamical importance and have fewer degrees of freedom
\cite{vlad}.

In this paper, we use numerical simulations to study the threshold
scalings in a weakly unstable collisionless Vlasov-Poisson system.
Depending on the sign of the Poisson equation this set of
equations describes either a plasma system or a gravitating
system. The saturation mechanism in a collisionless gravitating
system is the same as for the bump-on-tail instability in plasmas
and threshold corresponds to the condition $\gamma_L=0$ in both
cases. We show in Section \ref{bas_eq} that the eigenfrequency
contains only an imaginary part, and therefore is the simplest
model to study the threshold. Section \ref{resls} describes the
results of computations of the critical exponents and demonstrates
that the scaling laws describing saturation are the same for
plasma and gravitation. Section \ref{sym_law} addresses the
scaling transformations of the phase space and the scaling law,
which appears as a result of this symmetry. The exponent which
describes correlations are obtained in Section \ref{hyp_d}, where
Fisher's equality $\gamma=\nu(2-\eta)$ is also proved. In Section
\ref{rel_d_uc} we show that the criticality in the system is
described by the Dirac propagator for a fermionic field.  We
obtain hyperscaling laws and calculate upper critical
dimensionalities in Section \ref{hyp_slaw}.

\section{Basic Equations}{\label{bas_eq}}

The eigenfrequencies and eigenvectors of oscillations in a
Vlasov-Poisson system are given by dispersion relation
\begin{equation}
    \label{difl}
    \varepsilon[\omega({\bf k}),{\bf k}]=0~,
\end{equation}
where $\varepsilon$ is the permittivity (dielectric permittivity
in the plasma case). The boundary between stable and unstable
cases is determined by the condition
\begin{equation}\label{plasm}
    \mbox{Im}[\varepsilon(\omega,{\bf k})]=0~,
\end{equation}
\cite{ichi}. For the bump-on-tail instability condition
(\ref{plasm}) simplifies to, $\gamma_L \equiv
\mbox{Im}(\omega)=0$, and criticality is related to the zero of
the imaginary part of the eigenfrequency. Therefore we can employ
a model which does not contain the real part; i.e.,
$\mbox{Re}(\omega)=0$. The simplest is the one-dimensional
self-gravitating Vlasov-Poisson model which is described by the
equations
\begin{eqnarray}
\label{dst}
    &&\frac{\partial f}{\partial t}+v\frac{\partial
f}{\partial
x}-\frac{\partial\Phi}{\partial x}\frac{\partial f}{\partial v}=0, \\
\label{pot}
    &&\frac{\partial^{2}\Phi}{\partial x^{2}}=\int^{\infty}_
    {-\infty}f(x,v,t)\,dv-1,
\end{eqnarray}
where $f(x,v,t)$ is the distribution function, and $\Phi$ is the
gravitational potential. Boundary conditions are assumed to be
periodic in the $x$-direction.

For the eigenfunctions
\begin{equation}
    \label{eigen1}
    {\bf X}=\sum_{m=-\infty}^\infty {\bf X}_m\exp(ik_mx)~,
\end{equation}
where
\begin{equation}\label{eigen2}
    {\bf X}=[f(x,v,t),\Phi(x,t)]^T~,
\end{equation}
\begin{equation}\label{eigen3}
    {\bf X}_m=[f_m(v,t),\Phi_m(t)]^T~,
\end{equation}
are the spatial Fourier components, the superscript $T$ stands for
transpose, $k_m=2\pi m/L$ is the wavevector, and $L$ is the system
length we find
\begin{eqnarray}
\label{dful}
 \dot f_m+ik_m v f_m &+& i \sum_{m=m'+m''} \frac{1}{k_{m'}}
 \int^{\infty}_
    {-\infty} f_{m'}\,dv \nonumber \\
 & \times & \frac{\partial f_{m''}}{\partial v}=0,
\end{eqnarray}
or, explicitly for the components $m=\{0,1,2\}$ and for $L=2\pi$,
\begin{eqnarray}
\label{dcm0}
 & &\dot f_0 +i \frac{\partial}{\partial v}(\rho_1 f_{-1}
 - \rho_{-1} f_1) = 0, \\
 \label{dcm1}
 & &\dot f_1 +i v f_1 + i \frac{\partial}{\partial v}\left(\rho_1 f_0
 + \frac{1}{2} \rho_{2} f_{-1}-\rho_{-1} f_{2}\right)= 0, \\
 \label{dcm2}
 & &\dot f_2  +i~2 v f_2 + i\frac{\partial}{\partial v}
 \left(\frac{1}{2} \rho_2 f_0
 + \rho_{1} f_{1}\right)=0,
\end{eqnarray}
where
\begin{equation}\label{dens}
    \rho_m(t)=\int^{\infty}_{-\infty} f_m(v,t)dv~,
\end{equation}
is the Fourier component of density, and $f_{-1} = f^*_{1}$.

For a Maxwellian distribution
\begin{equation}
f_0(v)=\frac{1}{\sqrt{2\pi}\sigma}
\exp\left(-\frac{v^2}{2\sigma^2}\right) ,
\end{equation}
the dispersion relation is
\begin{equation}
\label{dr} 1+\frac{1}{2}\frac{\sigma^{2}_{J}(m)}{\sigma^{2}}
\left.\frac{dZ(\zeta)}{d\zeta}\right|_{\zeta=\zeta_m}=0,
\end{equation}
where $Z$ is the plasma dispersion function \cite{fr:com},
$\zeta_m=z_m/\sqrt{2}$, $z_m=\omega_m/(k_m\sigma)$, and $\sigma$
is velocity dispersion. In Eq. (\ref{dr}),
\begin{equation}
    \sigma^{2}_{J}(m)=\frac{1}{k^{2}_{m}}\equiv \frac{1}{m^2}
    \label{crvel}
\end{equation}
is the critical (Jeans) velocity dispersion for the mode $m$, and
$\rho_{0}$ is the background density.  For small $|z|\ll 1$ [i.e.,
$\sigma^{2}/\sigma^{2}_{J}(m)\ll 1$] the dispersion relation reads
\begin{equation}\label{drelat}
    \varepsilon(\omega_m,k_m)=1-\frac{\sigma^2_m}{\sigma^2}
    \left(1+i\sqrt{\frac{\pi}{2}}\frac{\omega_m}{k_m
    \sigma}\right)=0~,
\end{equation}
or
\begin{equation}
    \label{om}
    \omega_{m}=-i\sqrt\frac{2}{\pi}~ k_m \sigma
    \left[\frac{\sigma^{2}-\sigma^{2}_{J}(m)}{\sigma^{2}_{J}(m)}
\right].
\end{equation}
which is remarkably simpler than in plasma case, where the
bump--on--tail instability and Landau damping appear due to the
wave--particle resonance at the phase velocity of the wave
$v_{ph}=\omega_{pe}/k$. The frequency spectrum in the case of
gravitation does not contain a real part, so the resonance occurs
at $v=0$; i.e., in the main body of the particle distribution. For
all $m$, $\sigma^{2}_{J}(m)>\sigma^{2}_{J}(m+1)$, thus, if we
write $\sigma^{2}_{J}(1)\equiv \sigma^{2}_{cr}$, the distance from
the instability threshold is
\begin{equation}\label{theta}
    \theta=\frac{\sigma^2-\sigma_{cr}^{2}}{\sigma_{cr}^{2}}~,
\end{equation}
analogously to $\epsilon=(T-T_{cr})/T_{cr}$. Using (\ref{om}) and
(\ref{theta}) the dispersion relation can be rewritten for
$\omega_1$ as
\begin{equation}\label{omsm}
    \omega_1=-i\sqrt{\frac{2}{\pi}}~\theta ~.
\end{equation}
Time is measured in the units of the free-fall time
$t_{dyn}=(\sqrt{G\rho_0})^{-1}$. Since $4\pi G=1$ and the density
$\rho_0=m_p n=1$, where $m_p$ is the particle mass, and $n$ is the
concentration, $t_{dyn}=2\sqrt{\pi}$ in the units assumed here.

\begin{figure}[t]
\includegraphics[width=8.6cm,clip,height=4cm]{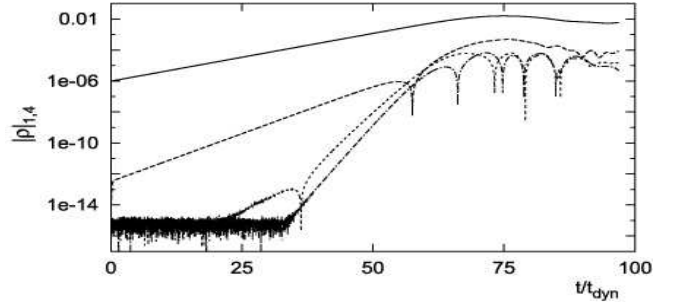}
\caption{Amplitudes of the first four harmonics, $m=1,2,3,4$. The
saturated level $A_{sat}$ is shown by the horizontal segment.}
\label{f1}
\end{figure}

\section{Results}{\label{resls}}

Dispersion relation (\ref{om}) shows that there are no unstable
modes above the threshold $\sigma^2_{cr}$,
$\sigma^2>\sigma^2_{cr}$, and therefore the system remains
invariant with respect to translations $x^\prime \to x+ \tau$,
where $\tau$ is any number. Below the threshold the mode $m=1$
becomes unstable, and therefore the $\it continuous$ symmetry
breaks and reduces to a lower {\it discrete} one with respect to
translations $x^\prime \to x+ L$. Therefore $\sigma^2_{cr}$ can be
considered as the critical point of a second-order phase
transition, and the amplitude of the mode $m=1$ as the order
parameter -- following the definition of Landau \cite{land}.

\begin{figure}[t]
\includegraphics[width=8.6cm,clip,height=4cm]{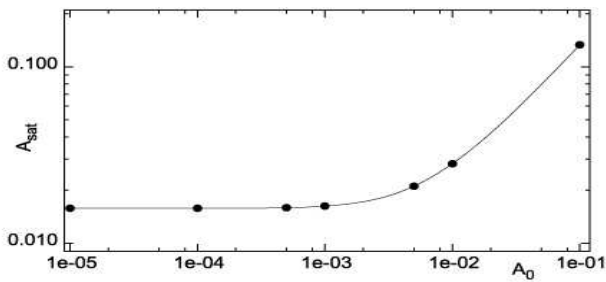}
\caption{The saturated amplitude $A_{sat}$ vs. $A_0$ for
$\theta=-0.05$. Circles are calculated values, the curve is a
spline approximation.} \label{f2}
\end{figure}

\subsection{Order parameter scaling}{\label{res_bt}}

Equations (\ref{dst}) and (\ref{pot}) with initial distribution
\begin{equation}\label{init}
    f(x,v,0)=f_0(v)[1+A_0\cos(k_1 x)],
\end{equation}
were integrated numerically using the Cheng-Knorr method
\cite{che:kn}. The amplitudes $A_m(t)=|\rho_m(t)|$ for
$m={1,2,3,4}$ are shown in Figure~\ref{f1}. The perturbation $m=1$
grows exponentially with the growth rate predicted by dispersion
relation (\ref{om}). Then, the growth saturates at some moment
$t=t_{sat}$ at the amplitude $A_{sat}=A_1(t_{sat})$.

Figure~\ref{f1} also shows the (exponential) growth of
perturbations with $m>1$ while (\ref{om}) predicts exponential
damping for these modes. This growth occurs because of nonlinear
coupling between modes since the term $\rho_1 f_1$ dominates over
the term $\rho_2 f_0$ in equation (\ref{dcm2}) initially, when
$f_2 \ll f_1$ and therefore one has $\gamma_2 = 2\omega_1$ for the
growth rate $\gamma_2$.

Figure~\ref{f2} shows that $A_{sat}$ is independent on $A_0$ for
small $A_0$, but there exists some threshold value of the initial
perturbation $A_0$, when becomes dependent on $A_0$. This
threshold amplitude $A_{thr}$ corresponds to the trapping
frequency $\omega_{b}=\sqrt{A_{thr}} \approx \omega_1$. At
$\omega_{b} \approx \omega_1$ the processes due to trapping become
as important as of the resonance between wave and particles
responsible for the linear Landau damping (or growth) in
collisionless media. Therefore to rule out the influence of
trapping processes on linear growth the amplitude $A_0$ must be
small to provide
\begin{equation}\label{ombom1}
 \omega_{b} \ll \omega_1.
\end{equation}

\begin{figure}[t]
\includegraphics[width=8.6cm,height=4cm]{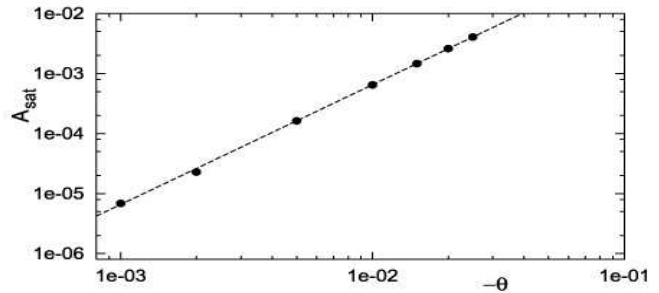}
\caption{Amplitude $A_{sat}$ as a function of $\theta$. Circles
represent calculated data, the dashed line is the power law best
fit.} \label{f3}
\end{figure}

\begin{figure}[t]
\includegraphics[width=8.6cm,height=4.5cm]{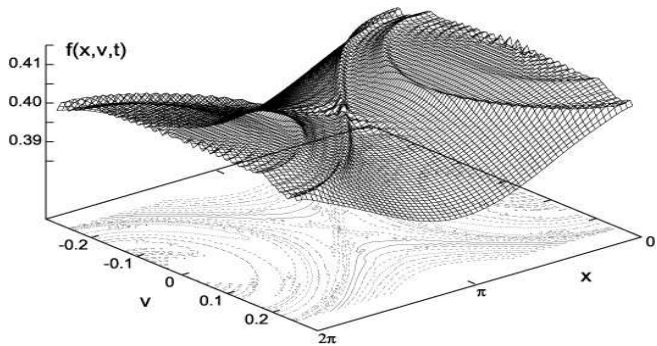}
\caption{Distribution function (surface) and its isocontours
(lines in the $x-v$ plane) in configuration space at the moment of
saturation $t=t_{sat}$.} \label{f6}
\end{figure}

The distribution function $f(x,v,t)$ is plotted in Figure~\ref{f6}
as a surface at the moment $t=t_{sat}$. Note that the distribution
function $f(x,v,t)$ becomes flat in the part of the $x-v$ domain
\begin{equation}\label{sepx}
    \frac{v^2}{2}+A_{sat}\cos(k_1x)\le A_{sat}
\end{equation}
separatrix, as predicted for the bump--on--tail instability
\cite{mazitov,oneil}. Outside this area the Fourier component
$f_1(v,t)$ remains modulated by the background Maxwellian
distribution as assumed at $t=0$, and the components with $m>1$
remain negligible \cite{iva}, so the dynamically important area
lies at $v\le|v_{sep}|$, where $v_{sep}=\pm 2\sqrt{A_{sat}}$. The
width of the dynamically important area must be small compared to
$\sigma$ at maximum amplitude (i.e., $A_{sat}$, $v_{sep}\ll
\sigma$), otherwise the background distribution will be altered by
evolution.

Assuming the above two criteria, $A_{sat}$ is calculated as a
function of $\theta$ and plotted in Figure \ref{f3}. From Figure
\ref{f3} we see that this dependence can be approximated by the
power law
\begin{equation}
\label{orpar} A_{sat}\propto (-\theta)^\beta,
\end{equation}
while $\beta=1.9950\pm 0.0034$ for $\theta\le 0$. Rewritten in
terms of the bounce frequency $\omega_b$ and the linear growth
rate $\gamma_L\equiv \mbox{Im}(\omega_1)$ the power law
(\ref{orpar}) becomes
\begin{equation}
\label{orpaw} \omega_{b} = c \gamma_L^\beta~,
\end{equation}
and the coefficient $c=3.22\pm 0.01$. These values are almost
identical to $\beta=2$, and to the coefficient $3.2$ in relation
(\ref{scalb}) as predicted and calculated for the bump--on--tail
instability \cite{oneil,crawf,candy}.

\subsection{Response scaling}{\label{res_gm_dl}}

\begin{figure}[t]
\includegraphics[width=8.6cm,height=4cm]{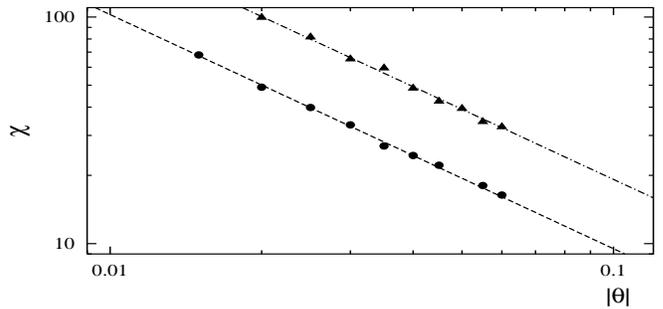}
\caption{Plot of $\chi$ vs. $|\theta|$. Circles ($\theta>0$) and
triangles ($\theta<0$) represent experimental data, the dashed
lines are the power--law best fits.} \label{f4}
\end{figure}

Subjecting the system to an external pump of the form
$F(x)=F_m\cos(k_mx+\varphi)$ allows one to calculate the other two
critical exponents, $\gamma$ and $\delta$, which describe the
response properties. The index $\gamma$ describes the divergence
of the susceptibility, which can be written as
\begin{equation}\label{susc}
    \chi(\theta)=\left.
     \frac{\partial A_{sat}(\theta)}{\partial F_1}
     \right|_{F_1 \to 0}~,
\end{equation}
for $m=1$. The results are shown by triangles ($\theta<0$) and
circles ($\theta>0$) in Figure~\ref{f4}.

Computation of $\chi$ at some $\theta$ requires at least 5 values
of $A_{sat}$ corresponding to the given $F_1$. At the same time,
$F_1$ must small enough to avoid the effects of $A_{sat}$
depending nonlinearly on $F_1$. Again it requires extensive
calculation of all quantities to high accuracy. In both cases
$\chi(\theta)$ is approximated by
\begin{equation}
\label{chi-g}
    \chi_{\pm}\propto|\theta|^{-\gamma_\pm},
\end{equation}
and $\gamma_-=1.028\pm 0.025$ for $\theta<0$, $\gamma_+=1.033\pm
0.016$ for $\theta>0$ giving
$\gamma_-\approx\gamma_+=\gamma\approx 1$. These exponents are
very close to the corresponding results for \cite{iva}, because
$\chi$ is the only linear coefficient, and this is common to both
wave-particle and wave-wave interactions.

The exponent $\gamma$ is the same as for the mean-field
thermodynamic models but, opposite to thermodynamics, the response
is {\it stronger} at $\theta<0$ than at $\theta>0$, as
Figure~\ref{f4} shows. The susceptibilities are
\begin{equation}\label{coef}
    \chi_-\approx 2 \chi_+ .
\end{equation}
This difference, as well as the appearance of scaling
(\ref{scalb}) and (\ref{orpaw}) with $\beta=2$ instead of
$\beta=1/2$ can be explained if one takes into account the
difference between the Landau-Ginzburg Hamiltonian
\begin{equation}\label{LGW}
    {\cal H}_{LG}=\frac{\kappa^2}{2}|\nabla\phi|^2
+\frac{\mu^2}{2}|{\phi}|^2+\frac{\lambda}{4!}(|\phi|^2)^2 ,
\end{equation}
[where $\phi$ is the order parameter and
$\mu^2\sim(T-T_{cr})/T_{cr}$], which describes the Ising
universality class, and the equation
\begin{equation}\label{craw}
\dot y= y \left [\gamma_L-\frac{1}{4\gamma_L^3} y^2 +
\mathcal{O}(y^4) \right ]~,
\end{equation}
which describes the amplitude of a weakly-unstable perturbation in
a one-species Vlasov-Poisson system \cite{crawf}. According to
(\ref{craw}) the maximum amplitude $y_{sat}$ at $\dot{y}=0$ scales
with $\gamma_L$ as $y_{sat}=\gamma_L^2$.

On the assumption that at $\gamma_L \lesssim 0$ the system
responds linearly to an external pump $\partial F$, one can obtain
the response $\partial y$ to $\partial F$ as
\begin{equation}\label{reps}
\gamma_L \partial y-\frac{3y_{sat}^2}{4\gamma_L^3} \partial y +
\partial F =0.
\end{equation}
However, equation (\ref{craw}) is not valid at $\gamma_L<0$ since
it predicts unlimited growth instead of damping in this case. At
small initial perturbation $y_0$ the correct evolution is given by
the linear equation $\dot{y}=\gamma_L y$, and the susceptibility
$\chi=\partial y/\partial F$ is
\begin{equation}\label{sl}
    \chi_+=\gamma_L^{-1} ,
\end{equation}
and at $\gamma_L>0$
\begin{equation}\label{sb}
    \chi_-=2\gamma_L^{-1} .
\end{equation}

\begin{figure}[t]
\includegraphics[width=8.6cm,clip,height=4cm]{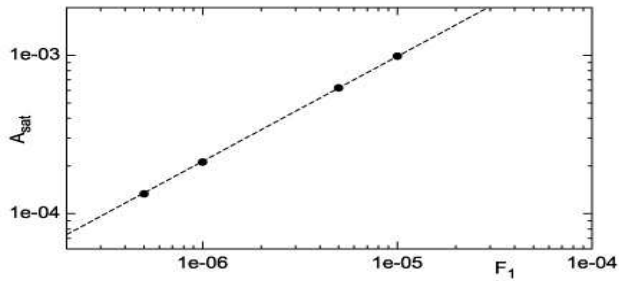}
\caption{$A_{sat}$ as a function of $F_1$. Circles represent
experimental data, the dashed lines are the power-law best fit.}
\label{f5}
\end{figure}

At the critical point $\theta=0$ (or $\gamma_L=0$) the response is
described by another critical exponent $\delta$
\begin{equation}
\label{delta-l}
A_{sat}\propto F_1^{1/\delta}.
\end{equation}
The results of simulation are plotted in Figure~\ref{f5}, giving
$\delta=1.503\pm 0.005$. This exponent cannot be obtained by the
previous simple assumption from (\ref{craw}) because of its
singularity at $\gamma_L=0$.

\section{Scaling laws and symmetries of the model}{\label{sym_law}}

The remarkable property of the critical exponents $\gamma$,
$\beta$, and $\delta$ is that they satisfy the Widom equality
(\ref{wideq}) \cite{widom} with high accuracy. In thermodynamics
the Widom equality is a consequence of the scaling of the Gibbs
free energy under the transformation
\begin{equation}
    \label{gibsym}
    \mathfrak{G}(\lambda^{a_{\epsilon}}\epsilon,\lambda^{a_B}B)=\lambda
     \mathfrak{G}(\epsilon,B)~,
\end{equation}
from which it can be derived straightforwardly \cite{st}. The
functions which comply the condition (\ref{gibsym}) are called
{\it generalized homogeneous functions}, and the condition itself
is termed a {\it homogeneity} condition.

The nature of this scaling for the marginally-stable
Vlasov-Poisson system is clear from Figure~\ref{f6}, where the
distribution function $f(x,v,t)$ is plotted at the moment
$t=t_{sat}$. The remarkable property of the critical dynamics is
the topological equivalence of the phase portraits for different
$\theta$: at the moments of saturation $t_1$ and $t_2$
corresponding to $\theta_1$ and $\theta_2$ we can write
\begin{equation}\label{c1}
    f(x,\lambda^{a_v}v,\theta_1,t_1)=\lambda f(x,v,\theta_2,t_2)~,
\end{equation}
or
\begin{equation}\label{c2}
    f_m(\lambda^{a_v}v,\theta_1,t_1)=\lambda f_m(v,\theta_2,t_2)~,
\end{equation}
for the Fourier component. Transformation between $t_1$,
$\theta_1$ and $t_2$, $\theta_2$ can be written as $t'\to
\lambda^{a_t}t$, $\theta'\to \lambda^{a_\theta}\theta$, so
\begin{equation}\label{c3}
    f_m(\lambda^{a_t}t,\lambda^{a_v}v,\lambda^{a_\theta}\theta)=\lambda
f_m(t,v,\theta).
\end{equation}

A weak external pump in the form $F(x)=F_1\cos(k_1 x+\varphi)$
creates a similar topology in the phase space because of the same
mechanism of the saturation and adds an additional variable to the
distribution function. The transformation can be written as
$F_1'\to \lambda^{a_{F_1}}F_1$. Finally for
$f_m=f_m(t,v,\theta,F_1)$ we can write the homogeneity condition
as
\begin{equation}
\label{hom}
    f_m(\lambda^{a_t}t,\lambda^{a_v} v,
    \lambda^{a_\theta}\theta,\lambda^{a_{F_1}}F_1)=
    \lambda f_m(t,v,\theta,F_1).
\end{equation}

The critical exponents $\beta$, $\gamma$, and $\delta$ can be
expressed via the scaling exponents $a_v$, $a_{\theta}$, and
$a_{F_1}$ from which the Widom equality for the Vlasov-Poisson
system can be proved directly (Appendix \ref{appn1}). They also
provide a deep insight into symmetry properties of the system.
According to expression (\ref{beta_a})
\[
    \beta=\frac{1+a_v}{a_\theta}~,
\]
rescaling the parameter of $\theta$ (or growth rate) also rescales
the distribution function $f(t,v,x)$ in the $v$-direction. This
situation differs significantly from thermodynamics where
\begin{equation}
    \label{bet_th}
    \beta=\frac{1-a_B}{a_\epsilon}.
\end{equation}
This expression rescales the normalized distance from the critical
point with external field $B$.

Substituting $A_{sat}$ according to the power law (\ref{orpar})
for the order parameter to $v_{sep}^2=4A_{sat}\propto
(-\theta)^{\beta}$ one can obtain (assuming $a_v=1$)
\begin{equation}\label{vb}
    v_{sep}\propto(-\theta)^{1/a_\theta}~,
\end{equation}
from which the scaling exponent is $a_\theta=1$ for $\beta=2$
($a_\theta=4$ for $\beta=1/2$). Remarkably, the two different
processes -- the linear growth of an unstable perturbation due to
the resonant wave-particle interaction and the subsequent
nonlinear saturation of this process due to particle trapping are
interrelated.

While there is no thermodynamic equilibrium in the collisionless
system considered here, one can define the quantity which
describes the response of the system to external thermal
perturbation, just as the specific heat capacity describes the
response of a thermodynamic system to heat transfer, $C=\delta
Q/dT$. For the case, considered here
\begin{equation}\label{heat}
    C=\frac{\delta Q}{d\theta} \equiv \frac{dV}{d\theta},
\end{equation}
where $V$ is the potential energy of the system. To calculate the
specific heat capacity, $V_{sat}$ corresponding to $A_{sat}$ is
used. The critical exponent $\alpha$ can be calculated
straightforwardly from (\ref{heat}) and (\ref{orpar}). Because
perturbations $m>1$ are negligible for $|\theta|\ll 1$,
$V_{sat}\propto A_{sat}\Phi_{sat}$, where $\Phi_{sat}=-A_{sat}$;
i.e., $V_{sat}\propto -\,(-\,\theta)\,^{2\beta}, \quad \theta<0$,
and the heat capacity is given by
\begin{equation}
    \label{heatcdef}
 C\propto-(-\,\theta)\,^{-\alpha}~,
\end{equation}
where
\begin{equation}
    \label{alph}
    \alpha=-(2\beta-1).
\end{equation}
The scaling law (\ref{alph}) can be proven using the homogeneity
condition (\ref{hom}) (Appendix \ref{appn1}). Remarkably, the
critical exponent $\alpha$ does not depend on the sign of
Poisson's equation, and result is the same for the plasma case.

Unlike thermodynamics where the relation between exponents
$\beta$, $\gamma$, and $\alpha$ is given by Rushbrooke's equality,
$\alpha+2\beta+\gamma=2$, the scaling law (\ref{alph}) does not
contain the critical exponent $\gamma$. Nevertheless, the set of
critical exponents $\alpha = -2.990 \pm 0.006$, $\beta=1.995\pm
0.003$, and $\gamma=1.031 \pm 0.021$ satisfy Rushbrooke's equality
with high accuracy.

\section{Correlation exponents}{\label{hyp_d}}

The correlation function of fluctuations for the field
\begin{equation}\label{field}
    E=-\frac{\partial \Phi}{\partial x}
\end{equation}
can be found from the fluctuation-dissipation theorem
\cite{cal_welt} as
\begin{equation}\label{fdt}
    \langle E^2 \rangle_{\omega {\bf k}}=\frac{T}{2\pi \omega}
\frac{\mbox{Im}[\varepsilon(\omega,{\bf k})]}{\varepsilon^2}~,
\end{equation}
\cite{kubo}, where the permittivity $\varepsilon$ is given by
(\ref{drelat}). Relation (\ref{fdt}) can be integrated using the
Kramers-Kronig dispersion relations, and in the static limit
$\omega\to 0$ (\ref{fdt}) becomes
\begin{equation}\label{static}
    \langle E^2 \rangle_{k_m}=\frac{4\pi\sigma^2}{m_p k_B}
    \left[1-\frac{1}{\varepsilon(0,k_m)}\right]~,
\end{equation}
where $k_B$ is the Boltzmann constant and $m_p$ is the particle
mass. This equation can be rewritten as
\begin{equation}\label{correl}
    \langle E^2 \rangle_{k_m}=-\frac{4\pi}{m_p k_B}
    \frac{\sigma^2}{\theta_m}~,
\end{equation}
where
\begin{equation}\label{thetm}
\theta_m=\frac{\sigma^2-\sigma^2_J(m)}{\sigma^2_J(m)}.
\end{equation}

The susceptibility $\chi$ can be written in terms of $\langle E^2
\rangle_{k_1}$ as
\begin{equation}\label{suscc}
    \chi=\langle E^2 \rangle_{k_1}\propto \theta^{-\gamma},
\end{equation}
$\gamma=1$. The combination of $\sigma$ and $\omega_1$ gives the
characteristic length for the system from the dispersion relation
(\ref{om})
\begin{equation}\label{corle}
    \xi=2\pi\frac{\sigma}{\omega_1},
\end{equation}
or, in terms of $\theta$,
\begin{equation}\label{coorth}
    \xi\propto \theta^{-\nu}~,
\end{equation}
as $\theta \to 0$. Therefore the critical exponent that
characterizes the correlation length is $\nu=1$. The correlation
function $\langle E^2 \rangle_{k_1}$ can be rewritten in terms of
$k_\xi=\xi^{-1}$ as
\begin{equation}\label{correch}
    \langle E^2 \rangle_{k_1}\propto k_\xi^{2-\eta}~,
\end{equation}
where $\eta$ is another critical exponent which characterizes the
correlation function. On the other hand, using (\ref{coorth}) one
can rewrite this expression as
\begin{equation}\label{correch1}
    \langle E^2 \rangle_{k_1}\propto \theta^{-\nu(2-\eta)}~,
\end{equation}
and, taking into account (\ref{suscc}),
\begin{equation}\label{correc}
    \theta^{-\gamma} \sim \theta^{-\nu(2-\eta)}~,
\end{equation}
from which finally we obtain the equality
\begin{equation}
    \label{fisher}
    \gamma=\nu(2-\eta).
\end{equation}
The last equality is known as Fisher's equality and gives the last
critical exponent, $\eta=1$.

\section{Relation with other universality classes}{\label{rel_d_uc}}

The correlation function (\ref{correl}) looks rather
counterintuitive, since at $\theta_m>0$ (damping waves), one has
$\langle E^2 \rangle_{k_m}<0$, and the noise is {\it imaginary}.
Nevertheless, this unusual situation has an analog -- for
particle-particle annihilation reactions of the type $Y+Y\to 0$
(corresponds to equation $dn/dt=-an^2,~a>0$), $Y\to 0$
($dn/dt=-an$) the correlation function is also negative because of
{\it anticorrelation} of particles \cite{hov_tau}. In the case
$\theta_m>0$ the amplitude $A_m \to 0$ as $t \to 0$. It is also
shown that the criticality due to these annihilation processes
belongs to a certain universality class which is different from
the Ising universality class \cite{hov_tau,lee} and therefore is
not described by the Landau-Ginzburg Hamiltonian (\ref{LGW}).

Another unusual quantity is the correlation length $\xi$ and the
wavevector $k_\xi=\xi^{-1}$, whose use allows us to establish the
validity of Fisher's equality for the collisionless system,
studied here. It is not related to the size of the system $L$ but
to the $\it fluctuations$ in the system which determine an average
path of correlated motion of particle in presence of these
fluctuations. As the system approaches the threshold, fluctuations
become correlated since the characteristic time of correlations
$\omega^{-1}\sim \theta^{-1}$ diverges as $\theta \to 0$. This
behavior is analogous to thermodynamic systems where the
correlation length is the only relevant scale near the critical
point as $\epsilon \to 0$.

To demonstrate that the criticality in the Vlasov-Poisson system
belongs to a different class, let us compare the critical
exponents corresponding to the Jeans instability in a
self--gravitating {\it hydrodynamical} system \cite{jeans}, using
the same approach. The dispersion relation for this system is
\begin{equation}
    \label{jeansf}
    \omega_m^2=c_s^2 k_m^2 -4\pi G\rho_0~,
\end{equation}
or
\begin{equation}
    \label{jeansr}
    \omega_m^2=(c_s^2-c^2_m)k_m^2~,
\end{equation}
where $c^2_m=4\pi G\rho/k_m^2$ is the critical velocity of sound,
corresponding to $\omega_m^2=0$. As for kinetic case if
$c^2>c_1^2=c^2_{cr}$, there are no unstable modes, and the
correlation length is
\begin{equation}
    \label{corrhyd}
    \xi_h=\frac{2\pi c_s}{\omega_1}\sim \frac{1}{k_1}\theta_f^{-1/2},
\end{equation}
where $\theta_f=(c_s^2-c^2_{cr})/c^2_{cr}$ is the reduced sound
velocity in a fluid. Here we have the mean-field exponent
$\nu_f=1/2$.

Assuming $m=1$ and dividing the both sides of dispersion relation
(\ref{jeansr}) on $c^2_s$ one can obtain the correlation function
as
\begin{equation}
\label{corh}
    G^{(2)}_h(k_{\xi_h},\theta_f)=\left(\frac{k^2_{\xi_h}}{k^2_1}
    -\theta_f\right)^{-1}~.
\end{equation}
This is the propagator of Euclidean theory or of the scalar boson
field \cite{amit} from which the Landau mean-field theory follows
automatically.

On the other hand dispersion relation (\ref{om}) for the
collisionless case gives
\begin{equation}
    \label{corcl}
    G^{(2)}(k_{\xi},\theta)=\left(i\sqrt{\frac{2}{\pi}}\frac{k_{\xi}}{k_1}
    -\theta \right)^{-1}~.
\end{equation}
For collisionless systems the propagator thus corresponds to the
vector {\it fermionic} field and describes a {\it different} class
of critical phenomena. In the language of quantum field theory the
parameters $\theta_f$ and $\theta$ are {\it bare masses}. Since
\begin{equation}
    \label{corcg}
    G^{(2)}(k,0) \propto \frac{1}{k^{2-\eta}}~,
\end{equation}
from (\ref{corh}) and (\ref{corcl}) one can obtain $\eta=0$ for
the case of hydrodynamics and $\eta=1$ for collisionless system.

\section{Hyperscaling laws}{\label{hyp_slaw}}

The approach assumed in the previous section allows us to
establish the hyperscaling law for the Vlasov-Poisson system which
involves the dimensionality $d$ along with critical exponents like
Josephson's law (\ref{josephl}). Using propagator (\ref{corcl})
which is the potential energy, the specific heat capacity $C$ in
$d$-dimensional space at $\theta \to 0$ can be obtained as
\begin{equation}
    \label{heatcf}
    C \sim \frac{\partial}{\partial \theta} \int d^d k_\xi
    G^{(2)}(k_{\xi},\theta)~.
\end{equation}
which gives
\begin{equation}
    \label{hcf}
    C \propto \xi^{2-d}~.
\end{equation}
With relation (\ref{coorth}), (\ref{hcf}) becomes
\begin{equation}
    \label{hcf1}
    C \propto \theta^{-\nu(2-d)}~.
\end{equation}
Taking into account the scaling law (\ref{heatcdef}) for the
specific heat capacity $C$ one can obtain the hyperscaling
relation which interrelates the exponents $\alpha$, $\nu$, and the
dimensionality $d$
\begin{equation}
    \label{ivanov_h}
    \alpha=\nu(2-d)~.
\end{equation}

The last equality reveals $d=2$ as the {\it upper critical
dimensionality} for the Vlasov-Poisson system since the heat
capacity becomes divergent if $d<2$, thus indicating the
importance of fluctuations in the critical area. It also shows
that the dimensionality corresponding to the critical exponents
$\alpha=-3$ and $\nu=1$ is $d=5$, fluctuations at $\theta \approx
0$ are insignificant, and therefore $\alpha=-3$, $\beta=2$,
$\gamma=1$, $\nu=1$, and $\eta=1$ are the mean-field exponents.

The use of the scalar field propagator (\ref{corh}) instead of
(\ref{corcl}) gives
\begin{equation}
    \label{ivanov_h1}
    \alpha=\nu(4-d)~,
\end{equation}
and at $\alpha=0$ the upper critical dimensionality is $d_c=4$
which is the Landau mean-field theory case for the Ising
universality class. However, relation (\ref{ivanov_h1}) is not
valid for the Vlasov-Poisson system because of its different
propagator. On the contrary to relations (\ref{ivanov_h}) and
(\ref{ivanov_h1}) which are valid for specific propagators
(\ref{corh}) and (\ref{corcl}), Josephson's law (\ref{josephl}) is
universal for all cases considered. With exponents $\nu=1$ and
$\nu_f=1/2$ it gives $d_c=2$ and $d_c=4$ as the upper critical
dimensionalities for the collisionless and hydrodynamic cases,
respectively, and $d=5$ for the exponents of the Vlasov-Poisson
system calculated here. Without going into details here, we note
that this universality appears because  the fundamental
description is given by the same functional integrals in both
cases. In particular for the free scalar bosonic field (no
interactions) the partition function is
\[
Z_G=\int\mathcal{D}\phi~\exp\left[-\int d^d{\bf x} {\cal H}_{0}
\right]~,
\]
where ${\cal H}_{0}$ is the Landau-Ginzburg Hamiltonian ${\cal
H}_{LG}$ (\ref{LGW}) without quadratic term. In the fermionic case
the Lagrangian for a Dirac spinor field is used instead of ${\cal
H}_0$.

\section{Conclusions}\label{conclus}

We have studied numerically and analytically a model
Vlasov--Poisson system near the point of a marginal stability. The
most important finding is that the criticality of the
Vlasov--Poisson model studied here belongs to a universality class
described by the propagator corresponding to a {\it fermionic
vector} field. This finding is in striking contrast with the
previous critical phenomena studies concerning systems whose
criticality belongs to universality classes corresponding to the
scalar {\it bosonic} fields, like the Ising universality class.

This fundamental discrepancy emerges from the {\it qualitative}
difference between objects considered: the Landau--Ginzburg
Hamiltonian (\ref{LGW}) takes into account spatial variations of
the order parameter via the local differential operator $\nabla$,
whereas the integro--differential operator for the Vlasov--Poisson
model acts on the distribution function containing the additional
dimension of velocity.

We have calculated numerically the critical exponents which
describe the critical state of the model and established
analytically that these exponents and the dimensionality are
interrelated by the scaling and hyperscaling laws like the Widom,
Rushbrooke, and Josephson laws at the formal dimensionality $d=5$.
The upper critical dimensionality is $d_c=2$ and since $d>d_c$ the
calculated exponents are the mean-field exponents, different from
those which one might expect the Landau--Weiss set of critical
exponents corresponding to the Ising mean--field model where
$d_c=4$. This is related to the higher dimensionality of the
Vlasov--Poisson kinetic problem associated with the velocity space
and to the type of the criticality of the Vlasov-Poisson systems,
which belongs to a universality class {\it different} from the
Ising universality class.

The critical exponents we have found here are $\alpha=-3$,
$\beta=2$, $\gamma=1$, $\delta=1.5$, $\nu=1$ and $\eta=1$. The
difference between this set and the set $\alpha \approx -2.814$,
$\beta \approx 1.907$, $\gamma \approx 1$, $\delta \approx 1.544$,
$\nu=1$ and $\eta=1$ \cite{iva} is because $A_{sat}$ is about 50
times larger for the latter case, thus causing {\it wave-wave}
interactions to dominate, thereby yielding a different
universality class. More important, the later exponents satisfy
scaling laws at {\it fractal} dimension $d \approx 4.68$
indicating {\it reduced} dimensionality because wave-wave
interactions have fewer degrees of freedom than wave-particle ones
\cite{vlad}.

\acknowledgments

This work was supported by the Australian Research Council and a
University of Sydney SESQUI grant.

\appendix

\section{Relation between the scaling and critical exponents}{\label{appn1}}

From the homogeneity condition (\ref{hom})
\begin{eqnarray}
\label{hom1a}
    f_m(\lambda^{a_t}t,~\lambda^{a_v} v,
    ~\lambda^{a_\theta}\theta,~\lambda^{a_{F_1}}F_1) \\ \nonumber
    =\lambda f_m(t,v,\theta,F_1)~,
\end{eqnarray}
for $\rho_m$ components by integration over $v$, one has
\begin{eqnarray}
\label{hom2a}
\lambda^{-a_v}\rho_m(\lambda^{a_t}t,~\lambda^{a_\theta}\theta,
~\lambda^{a_{F_1}}F_1) \\ \nonumber =\lambda\rho_m(t,\theta, F_1).
\end{eqnarray}
For any two $A_{sat}=\rho_1(t_{sat})$ and
$A^\prime_{sat}=\rho_1(t^\prime_{sat})$ one can write
\begin{equation}
\label{hom4a}
\lambda^{-a_v}A_{sat}(\lambda^{a_\theta}\theta,~\lambda^{a_{F_1}}F_1)=
\lambda A_{sat}(\theta,F_1).
\end{equation}

Assuming $\lambda=(-1/\theta)^{1/a_{\theta}}$, the critical
exponent $\beta$ can be rewritten in terms of the scaling
exponents $a_v$ and $a_\theta$ as
\begin{equation}
    \label{beta_a}
    \beta=\frac{1+a_v}{a_\theta}~.
\end{equation}
In the similar way for $\gamma$ and $\delta$ one can write
\begin{equation}
    \label{gam_p}
    \gamma=\frac{-a_v-1+a_{F_1}}{a_\theta},
\end{equation}
\begin{equation}
    \label{delta_a}
    \delta=\frac{a_{F_1}}{1+a_v}~,
\end{equation}
and the Widom relation follows from (\ref{gam_p})
straightforwardly:
\begin{eqnarray}
    \label{equal_pw}
    \gamma=\frac{-a_v-1+a_{F_1}}{a_\theta}
    & = &  -\frac{1+a_v}{a_\theta}+\frac{a_{F_1}}{a_\theta} \nonumber \\
    &=&-\beta+\beta\delta \nonumber \\
    &=& \beta(\delta-1)~.
\end{eqnarray}

Equations (\ref{beta_a})-(\ref{delta_a}) can be rewritten in
matrix form as
\begin{equation}
    \label{mfr}
    {\bf W}{\bf A}={\bf X}~,
\end{equation}
where ${\bf A}=[a_\theta,a_v,a_{F_1}]^T$, ${\bf
X}=[1,-1,-\delta]^T$, and the matrix ${\bf W}$ is
\begin{equation}
{\bf W}=\left (
\begin{array}{cccc}
\beta& -1 & 0\\
\gamma& 1 & -1\\
0& \delta & -1
\end{array}
\right) .
\end{equation}
The determinant of ${\bf W}$ is
\begin{equation}\label{detw}
    \det{\bf W}=-\beta+\delta \beta -\gamma \equiv 0~.
\end{equation}

Using (\ref{delta_a}) to eliminate $a_v$, the system (\ref{mfr})
can be reduced to
\begin{eqnarray}\label{s11}
    \beta a_\theta & - & \frac{1}{\delta}a_{F_1}=0~, \\
    \gamma a_\theta & + & \left(\frac{1}{\delta}-1\right)a_{F_1}=0~,
\end{eqnarray}
for which solution exists only if the Widom equality
$\gamma=\beta(\delta-1)$ holds. Therefore $a_\theta$ and $a_{F_1}$
can be formally considered as the eigenvectors of $\bf{W}$ whose
eigenvalue is $\lambda=0$. In particular
\begin{equation}\label{scF}
    a_\theta=\frac{1}{\beta+\gamma}~a_{F_1}
\end{equation}
which indicates that rescaling of the distribution function under
an external pump is equivalent to rescaling due to the field which
appears for nonzero order parameter.

\section{Rushbrooke's law for Vlasov--Poisson system}

The heat capacity can be formally defined as
\begin{equation}
C=\frac{\delta Q}{d\theta}\equiv \frac{dV}{d\theta},
\end{equation}
where $V$ is the potential energy of the system. To calculate the
specific heat capacity, $V_{sat}$ corresponding to $A_{sat}$ is
used.

Because perturbations $m>1$ are negligible for $|\theta|\ll 1$,
$V_{sat}\propto A_{sat}\Phi_{sat}$, where $\Phi_{sat}=-A_{sat}$,
and
\begin{equation}
V_{sat} \propto A_{sat}^2.
\end{equation}
From (\ref{hom4a}) one can obtain
\begin{equation}
    \label{cpheat}
    \frac{\partial}{\partial \theta}\lambda^{-2a_v}
    A^2_{sat} (\lambda^{a_\theta}\theta,~\lambda^{a_{F_1}}F_1)=
    \frac{\partial}{\partial \theta}\lambda^2 A^2_{sat}(\theta,F_1),
\end{equation}
or
\begin{equation}\label{cph1}
    \frac{\partial}{\partial \theta}\lambda^{-2
    a_v-2}A^2_{sat}(\lambda^{a_\theta}\theta,~\lambda^{a_{F_1}}F_1)=
    \frac{\partial}{\partial \theta}A^2_{sat}(\theta,F_1)~.
\end{equation}

Assuming $\lambda=\theta^{-1/a_{\theta}}$ and $F_1=0$, Equation
(\ref{cph1}) can be rewritten as
\begin{equation}\label{ww2}
    \frac{\partial}{\partial \theta}[\theta^{(2a_v+2)/a_\theta}A_{sat}^2(-1,0)]=
    \frac{\partial}{\partial \theta}A_{sat}^2(\theta,0),
\end{equation}
or
\begin{equation}\label{ww3}
    \frac{2a_v+2}{a_\theta}~A_{sat}^2(-1,0)~\theta^{(2a_v+2)/a_\theta-1}=
    \frac{\partial}{\partial \theta}A_{sat}^2(\theta,0),
\end{equation}
or
\begin{equation}\label{ww4}
    \frac{2a_v+2}{a_\theta}~A_{sat}^2(-1,0)~\theta^{\frac{(2a_v+2)}{a_\theta}-1}=
    C(\theta,0).
\end{equation}
Equation (\ref{ww2}) has the form of the power law,
$C(\theta,0)\propto \theta^{-\alpha}$, with
\begin{eqnarray}\label{alpha}
    \alpha=-2\frac{a_v+1}{a_\theta}+1=-2\beta+1.
\end{eqnarray}
The last relation corresponds to Rushbrooke's equality
$\alpha+2\beta+\gamma=2$ at $\gamma=1$.


\begin{thebibliography}{zz}
\bibitem{st}
H. E. Stanley, {\it Introduction to Phase Transitions and Critical
Phenomena} (Clarendon, Oxford, 1971).

\bibitem{land}
L. D. Landau and E. M. Lifshitz, {\it Statistical Physics} (Pergamon,
Oxford, 1980)

\bibitem{bump}
E. Frieman, S. Bodner, and P. Rutherford, Phys. Fluids {\bf 6},
1298 (1963).

\bibitem{fried}
B. C. Fried, C. S. Liu, R. W. Means, and R.Z. Sagdeev, Plasma
Physics Group Report PPG-93, University of California, Los
Angeles, 1971 (unpublished).

\bibitem{levin_ea}
M. B. Levin, M. G. Lyubarsky, I. N. Onishchenko, V. D. Shapiro,
and V. I. Shevchenko, Sov. Phys. JETP {\bf 35}, 898 (1972).

\bibitem{land1}
L. D. Landau, J. Phys. {\bf 10}, 25 (1946).

\bibitem{hopf}
J. E. Marsden and M. McCracken, {\it The Hopf Bifurcation and Its
Applications} (Springer-Verlag, New York, 1976).

\bibitem{latora}
V. Latora, A. Rapisarda, and S. Ruffo, \prl {\bf 80}, 692 (1998).

\bibitem{rozen}
A. Simon and M. Rosenbluth, Phys. Fluids {\bf 19}, 1567 (1976).

\bibitem{crawf}
J. D. Crawford, Phys. Plasmas {\bf 2}, 97 (1995).

\bibitem{denavit}
J. Denavit, Phys. of Fluids {\bf 28}, 2773 (1986).

\bibitem{candy}
J. Candy, J. Comp. Phys. {\bf 129}, 160 (1996).

\bibitem{amit}
D. J. Amit, {\it Field theory, the renormalization group and
critical phenomena} (World Scientific, Singapore, 1984).

\bibitem{widom}
B. Widom, \jcp {\bf41}, 1633 (1964).

\bibitem{joseph}
B. D. Josephson, Proc. Phys. Soc. {\bf 92}, 269, 276 (1967).

\bibitem{staufer}
D. Stauffer and A. Aharony, {\it Introduction to percolation
theory} (Taylor \& Francis, London, 1994).

\bibitem{domb}
C. Domb, Adv. Phys. {\bf 9}, 45 (1960).

\bibitem{gold}
N. Goldenfeld, {\it Lectures on phase transitions and the
renormalization group} (Addison-Wesley, Reading, Massachusetts,
1992).

\bibitem{iva}
A. V. Ivanov, \apj {\bf 550}, 622 (2001).

\bibitem{rob}
P. A. Robinson, Rev. Mod. Phys. {\bf 69}, 507, (1997).

\bibitem{vlad}
S. V. Vladimirov, V. N. Tsytovich, S. I. Popel, and F. Kh.
Khakimov, {\it Modulational Interactions in Plasmas} (Kluwer,
Dordrecht, 1995).

\bibitem{ichi}
S. Ichimaru, D. Pines, and N. Rostoker, \prl \textbf{8} 231
(1962).

\bibitem{fr:com}
B. D. Fried and S. D. Conte, {\it Plasma Dispersion Function: The
Hilbert Transform of the Gaussian} (Academic Press, New York, 1961).

\bibitem{che:kn}
C. Z. Cheng and G. Knorr, J. Comput. Phys. {\bf 22}, 330 (1976).

\bibitem{mazitov}
R. K. Mazitov, Zh. Prikl. Mekh. Fiz. {\bf 1}, 27 (1965).

\bibitem{oneil}
T. M. O'Neil, J. H. Winfrey, and J. H. Malmberg, Phys. Fluids {\bf 14},
1204 (1971).

\bibitem{jeans} J. Jeans,
{\it Astronomy and Cosmogony} (University Press, Cambridge, 1928).

\bibitem{cal_welt}
H. B. Callen and T.A. Welton, Phys. Rev. \textbf{83}, 34 (1951).

\bibitem{kubo}
R.J. Kubo, J. Phys. Soc. Japan {\bf 12}, 570 (1957).

\bibitem{hov_tau}
M. J. Howard and U. C. T\"auber, J. Phys. A: Math. Gen. {\bf 30},
7721 (1997).

\bibitem{lee}
B.P. Lee, J. Phys. A: Math. Gen. {\bf 27}, 2633 (1994).

\end{thebibliography}
\end{document}